\documentclass[fleqn,10pt]{wlscirep}
\title{Four-dimensional entanglement distribution over 100 km}

\author[1,*]{Takuya Ikuta}
\author[1]{Hiroki Takesue}
\affil[1]{NTT Basic Research Laboratories, NTT Corporation, 3-1 Morinosato Wakamiya, Atsugi, Kanagawa 243-0198, Japan}
\affil[*]{ikuta.takuya@lab.ntt.co.jp}

\usepackage{multirow}
\usepackage{color}
\usepackage{braket}
\newcommand{\op}[1]{\hat{ #1}}
\newcommand{\twobit}{\rm 2\mathchar`-bit}
\newcommand{\onebit}{\rm 1\mathchar`-bit}

\begin{abstract}
High-dimensional quantum entanglement can enrich the functionality of quantum information processing.
For example,
it can enhance the channel capacity for linear optic superdense coding and decrease the error rate threshold of quantum key distribution.
Long-distance distribution of a high-dimensional entanglement is essential for such advanced quantum communications over a communications network.
Here, we show a long-distance distribution of a four-dimensional entanglement.
We employ time-bin entanglement, which is suitable for a fibre transmission,
and implement scalable measurements for the high-dimensional entanglement using cascaded Mach-Zehnder interferometers.
We observe that a pair of time-bin entangled photons has more than 1 bit of secure information capacity over 100 km.
Our work constitutes an important step towards secure and dense quantum communications in a large Hilbert space.
\end{abstract}
\begin{document}


\flushbottom
\maketitle
%
%
\thispagestyle{empty}

\section*{Introduction}

Long-distance distribution of quantum entanglement \cite{Marcikic2004,Fedrizzi2009,Dynes2009,Inagaki2013,Cuevas2013, Yin2017} is
essential for quantum communications.
Distributed quantum entanglement enables two distant parties to perform communication protocols
that are impossible with classical information processing,
such as quantum teleportation \cite{Bennett1993}
and quantum key distributions \cite{Ekert1991}.
Quantum communications using distributed quantum entangled qubits has been demonstrated over distances longer than 100 km
\cite{Takesue2015,Ma2012,Yin2012,Ursin2007,Takesue2010},
which can cover a communications network in an urban area.
Quantum communications using entanglement has mainly focused on a two-dimensional bipartite state.
Currently, however,
high-dimensional entanglement---entangled qudits---is attracting much attention
because its larger Hilbert space allows us to enrich the functionality of quantum communications protocols.
Entangled qudits have been investigated on the basis of various optical orthogonal modes,
including orbital angular momentum (OAM) \cite{Dada2011a,Zajonc2012,Krenn2014,Agnew2011},
frequency \cite{Kues2017, Imany2017,Bernhard2013},
time \cite{Ansari2017,Ikuta2016a,Nowierski2016,Bessire2014,Thew2004,DeRiedmatten2004},
and combinations of different optical modes, or hyper entanglement \cite{Barreiro2005}.
Entangled qudits can be used to overcome the channel capacity limit for linear photonic superdense coding \cite{Barreiro2008}.
Furthermore, they enable us to perform high-dimensional quantum key distribution.
Using high-dimensional entanglement,
we can decrease the threshold of the symbol error rate
and increase the information capacity of a secure channel \cite{Cerf2002,Sheridan2010},
which has been demonstrated with a free-space optical setup in the laboratory \cite{Mafu2013}.
For such quantum communications over a communications network,
it is necessary to distribute high-dimensional entanglement.
Entangled qudits have been distributed over 1.2 and 15 km on free-space and fibre-based optical links, respectively \cite{Steinlechner2016,Jin2016a}.
However,
a remaining challenge is long-distance distribution that can cover a communications network in an urban area.

For long-distance distribution for advanced quantum communications,
it is important to generate maximally entangled qudits efficiently.
One way to do this is to exploit optical nonlinear effects.
In this process,
probability amplitudes of generated photons are often different from those of the maximally entangled state.
When the difference is not negligible,
a filtering process after the two-photon generation \cite{Dada2011a} 
or complicated preparation of the pump light \cite{Kovlakov2017} is required.
In particular,
filtering to generate the maximally entangled state reduces the generation rate of the entanglement,
leading to a longer measurement time.
Another problem is qudit degradation caused by various disturbances in a transmission channel.
Spatial-mode-based qudits are especially vulnerable to these disturbances.
OAM transmission over 143 km was attempted by using classical light \cite{Krenn2016}.
However,
it is necessary to build an active high-speed stabilization system composed of optical components for high-dimensional entanglement distribution.
Finally,
since a qudit essentially has many parameters characterizing the state,
measurements for qudit characterization are more complex than those for qubits.
In particular,
we cannot confirm for certain a high-dimensional entanglement with a single two-dimensional subspace measurement
even if the generated state is a high-dimensional entanglement \cite{DeRiedmatten2004, Dynes2009, Inagaki2013}.

Here we report four-dimensional entanglement distribution over 100 km of fibre.
We employ time-bin entangled qudits generated via spontaneous parametric down-conversion,
where the maximally entangled state is generated without extra filters \cite{Ikuta2016a,DeRiedmatten2004}.
The time-bin state is robust against disturbances in fibre transmission regardless of its dimension,
which contributes to the success of long-distance distribution.
Although fibre length variation in a long-time measurement leads to fluctuations in the photon detection times and disturbs the measurement \cite{Inagaki2013},
we realize a stable measurement by implementing an algorithm that automatically tracks fluctuations of photon detection times.
The state after fibre transmission is evaluated by using the quantum state tomography (QST) scheme proposed by the authors,
with which we can significantly reduce the complexity of the experimental procedure \cite{Ikuta2017}.
We show that the four-dimensional entanglement is conserved after 100-km distribution.
We also discuss the secure information capacity of the measured photon pair.
The results indicate that the measured photon pair has a secure information capacity of more than 1 bit even after the distribution over 100 km.

\section*{Results}

\subsection*{Experimental setup}

\begin{figure*}[hbtp]
	\includegraphics[width=17.0cm]{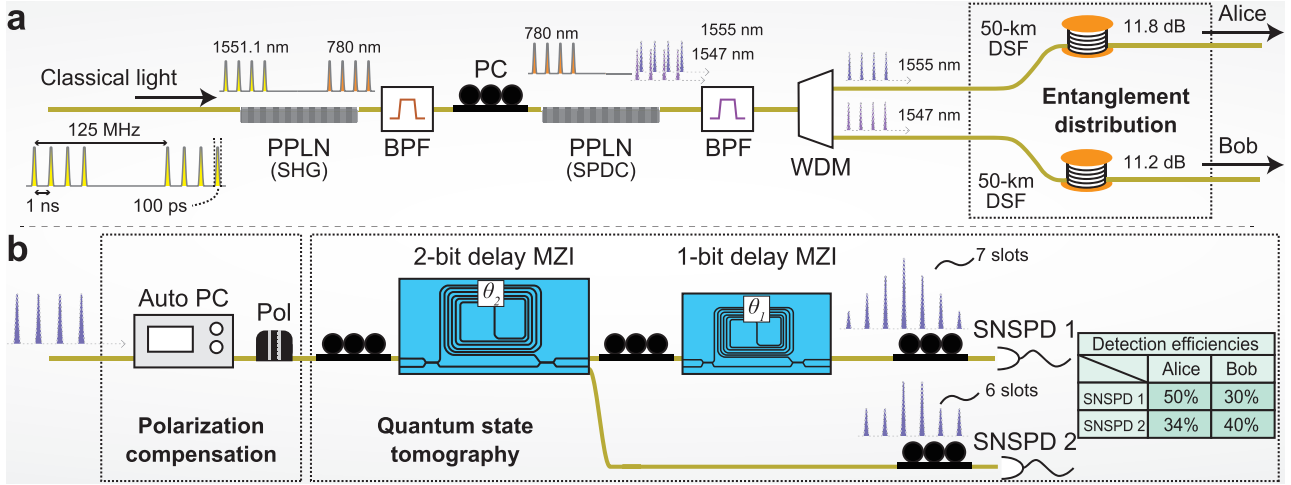}
	\caption{\label{fig:expsetup} {\bf Experimental setup.}
	({\bf a}) Generation and distribution of four-dimensional time-bin entanglement.
	({\bf b}) Alice and Bob's measurements.
	PPLN, periodically poled lithium niobate waveguide;
	BPF, band pass filter;
	PC, polarization controller;
	WDM, wavelength demultiplexing filter;
	DSF, dispersion shifted fibre spool;
	Auto PC, remote controllable polarization controller;
	Pol, polariser;
	MZI, Mazh-Zehnder interferometers;
	SNSPD, superconducting nanowire single photon detector.
	The inset shows the detection efficiencies of the SNSPDs.}
\end{figure*}

The setup for generation and distribution of the four-dimensional entanglement is shown in Fig. \ref{fig:expsetup}a.
We modulated the intensity of a continuous-wave laser light with a 1551.1-nm wavelength and 10-$\mu$s coherence time
to generate four sequential pulses.
The pulse duration, time interval, and repetition frequency were 100 ps, 1 ns, and 125 MHz, respectively.
These sequential pulses were launched into a periodically poled lithium niobate (PPLN) waveguide
to generate pump pulses through second harmonic generation (SHG).
The pump pulses were launched into another PPLN waveguide
to create a four-dimensional time-bin maximally entangled state via spontaneous parametric down-conversion (SPDC).
A time-bin entangled state generated via SPDC is given by
\begin{equation}
\ket{\psi} = \sum_{k=0}^{d-1} c_k \ket{k}_s \otimes \ket{k}_i,
\end{equation}
where $c_k$ is a probability amplitude satisfying $\sum |c_k|^2 = 1$,
$d$ is the number of pump pulses for SPDC.
$\ket{k}_s$ and $\ket{k}_i$ denote states where signal and idler photons exist in the $k$-th time slot, respectively.
By modulating the pump pulse in the corresponding time slot,
$c_k$ can be simply controlled without extra filters \cite{Ikuta2016a,DeRiedmatten2004}.
Here, we equalized the intensities of the four sequential pump pulses
to generate the maximally entangled state.
In our experimental setup,
the signal and idler photons had wavelengths in the telecommunications C-band,
where the photon loss in fibre transmission is minimized.
The signal and idler photons with 1555- and 1547-nm wavelengths, respectively,
were separated by a wavelength demultiplexing (WDM) filter and launched into 50-km optical fibre spools.
We used dispersion shifted fibres (DSFs) to avoid broadening the pulse widths of the photons.
The fibre spools for the signal and idler photons had 11.8- and 11.2-dB transmission losses, respectively.

After the distribution over the fibres,
the signal and idler photons were sent to two receivers, Alice and Bob.
Each receiver performed a measurement using the setup depicted in Fig. \ref{fig:expsetup}b.
Because our measurement setup had polarization dependence,
the receivers first compensated for the polarizations of the photons using remote controllable polarization controllers and polarisers
(see Supplementary Information.)
After polarization compensation,
the photons were launched into the measurement setup.
The measurement setup was composed of 1- and 2-bit delay Mach-Zehnder interferometers (MZIs)
and two superconducting nanowire single photon detectors (SNSPDs).
Each MZI had two input ports and two output ports.
One input port of the 1-bit delay MZI was connected to an output port of the 2-bit delay MZI.
SNSPD 1 and 2 were connected to an output port of the 1-bit delay MZI and the remaining output port of the 2-bit delay MZI, respectively.
The 1- and 2-bit delay MZIs had 1- and 2-ns delay times, respectively.
The phase differences between the short and long arms,
$\theta_1$ and $\theta_2$, of the 1- and 2-bit delay MZIs, respectively,
were set at either $0$ or $\pi/2$ for QST.
Because we employed a four-dimensional time-bin state and 1- and 2-bit delay MZIs,
the photon could be detected in seven and six different time slots at SNSPD 1 and 2, respectively (see Fig. \ref{fig:expsetup}b).
Depending on the detected time $t$,
the index of the SNSPD, $x$,
and phase differences $\theta_1$ and $\theta_2$,
the photon was observed by different measurement operators $\op{E}_{t x \theta_1 \theta_2}$ (see Methods).
By comparing the coincidence counts and measurement operators under all possible combinations of $(t, x, \theta_1, \theta_2)$ for Alice and Bob,
we reconstructed the density operator of the two photons, $\op{\rho}$ \cite{Ikuta2017}.
Note that 
we performed QST with only 16 measurement settings
because we only changed $\theta_1$ and $\theta_2$,
each of which could take two possible values.
This significant simplification of the measurement is possible
because different measurements are simultaneously implemented in a time-bin state measurement using delayed interferometers.

\subsection*{\label{subsec:stab}Tracking of the photon detection times}

\begin{figure}[hbtp]
	\includegraphics[width=8.5cm]{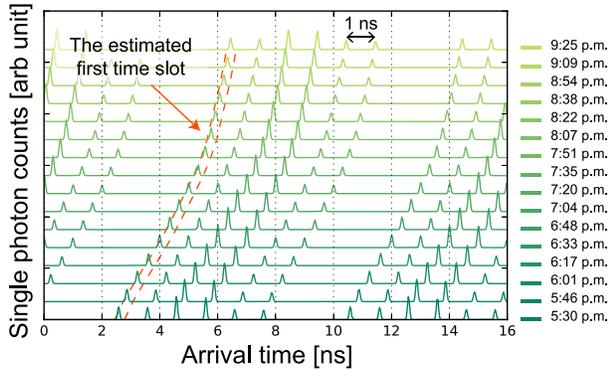}
	\caption{\label{fig:stabilize}{\bf Result of photon detection time tracking.}
	Histograms of single photon counts in long-time measurements at SNSPD 2 for Alice are shown.
	Single photon counts were accumulated in one minute for each histogram.
	The photon detection time drifted about 4 ns during the measurement,
	which was longer than the total duration of the four-dimensional time-bin state.
	}
\end{figure}

Here we describe our scheme for tracking the photon detection times.
For maximally entangled time-bin qubits,
we can track the fluctuation of the photon detection time by selecting the time slot that shows the highest single photon count in a histogram.
However,
this method is not valid for the present experiment because we have two time slots showing the highest single photon count at SNSPD 2.
To track the photon detection time precisely and deterministically,
we used the cross correlation function $g(\tau)$,
given by
\begin{equation}
g(\tau) = \int_{0}^{8T} h_{i}(t-\tau \bmod 8T)h_{m}(t ) dt ,
\end{equation}
where $h_i(t)$ and $h_m(t)$ are an ideal and a measured histogram of single photon counts, respectively,
and $T$ is the time interval between the time slots.
We employed $h_i(t) = \sum_{l=0}^{3} \sum_{k=0}^{3} \delta(t - kT - lT)$ for SNSPD 1
and
$h_i(t) = \sum_{l=0}^{1} \sum_{k=0}^{3} \delta(t - kT - 2lT)$ for SNSPD 2,
where $\delta(t)$ is the Dirac delta function.
This correlation function returns the highest value
when $\tau$ is equal to the position of the first time slot in the measured histogram of the single photon counts.
Therefore,
we can deterministically track the fluctuation of the photon detection time.
The measured histograms of the single photon counts at the SNSPD 2 for Alice
and the first time slots estimated by the cross correlation function with a 0.33-ns time window are shown in Fig. \ref{fig:stabilize}.
The estimated first time slot precisely overlapped the first time slot indicated by the measured histogram.
After this compensation,
we analysed coincidence counts between Alice and Bob.

\subsection*{Qualities of the reconstructed state}

We performed coincidence measurements for QST four times after the long-distance distribution.
The measurement time for each phase setting of the MZIs, $\theta_1$ and $\theta_2$, was 15 min;
thus, it took totally four hours to perform each QST.
The average number of photon pairs per qudit was 0.03.
The average single photon counts at SNSPD 1 and 2 for Alice (Bob) were 3.3 and 7.7 (2.9 and 12) kcps, respectively.
From the measured coincidence counts,
we reconstructed the density operator of the four-dimensional entanglement
by using maximum likelihood estimation \cite{Ikuta2017, James2001,Thew2002}.
It is known that a QST using maximum likelihood estimation leads to large systematic errors if the number of coincidence counts is small \cite{Schwemmer2015}.
In our experiment, the total coincidence counts per trial was sufficiently large ($>600,000$),
which means that such errors were expected to be small.
The reconstructed density operator is shown in Fig. \ref{fig:rho}.
All measured coincidence counts and reconstructed density operators are provided in Supplementary Data 1 and 2, respectively.
Both the real and imaginary parts show characteristics close to the four-dimensional maximally entangled state.
We also derived five figures of merit from the reconstructed density operators to quantify the quality of the two photon state after the distribution,
which are summarized in Table \ref{tab:figmerit} (see Methods for the definitions.)
The measured fidelity and trace distance were close to one and zero, respectively,
which indicated the reconstructed state was close to the four-dimensional maximally entangled state.
Moreover, the reconstructed state was close to a pure state
because the measured linear entropy and von Neumann entropy were low.
Furthermore,
conditional entropy ensured that the measured two photons were not a two-dimensional entanglement.
Note that conditional entropy cannot be negative without entanglement \cite{Horodecki1996,Horodecki1996a}.
In addition,
the minimum value of conditional entropy for a two-dimensional two-photon state is $-1$ bit.
We emphasize that
we obtained a conditional entropy of $-1.557$ bit,
which is smaller than the minimum value for two-dimensional entanglement by eight standard deviations.
These results indicate that the four-dimensional entanglement was conserved after the distribution over 100 km.

\begin{figure*}[hbtp]
	\includegraphics[width=17.0cm]{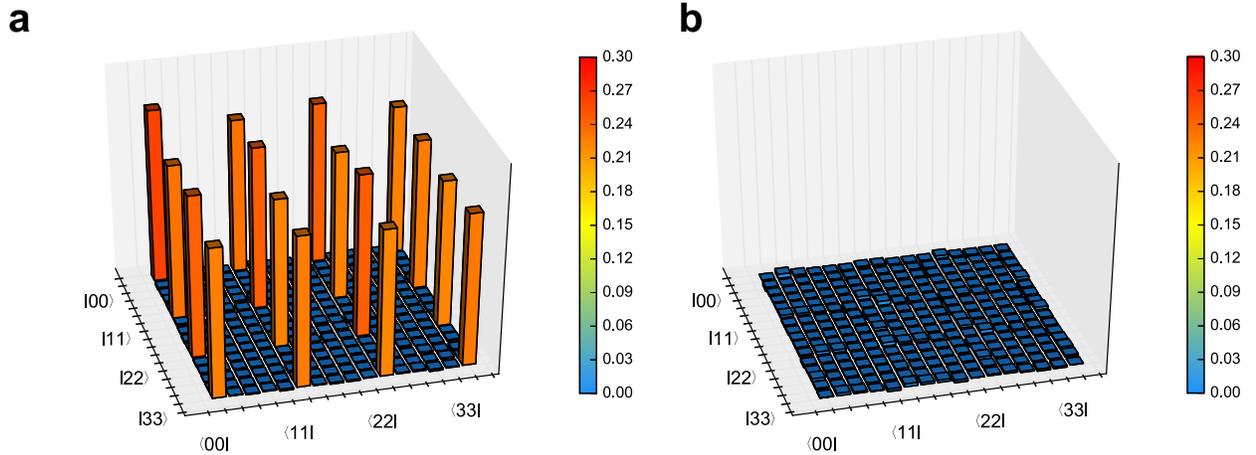}
	\caption{\label{fig:rho}{\bf Experimental results.}
	({\bf a}) Real and ({\bf b}) imaginary parts of the density operator reconstructed by QST.
	The data shown here were averaged over four trials.
	To increase readability,
	the local phase rotation $\op{U}(\phi) = \sum_k \exp(-ik\phi) \ket{k}_s \bra{k}_s$ was multiplied after averaging $\op{\rho}$,
	where $\phi$ was calculated from the probability amplitudes of $\ket{00}$ and $\ket{11}$ in the eigenstate with the largest eigenvalue.
	}
\end{figure*}

\begin{table}
\caption{\label{tab:figmerit}
	{\bf Figures of merit for the reconstructed density operators.}
	The errors were estimated as standard deviations in the four experimental trials,
	which means that
	the errors included not only statistical characteristics of the photon counts but also experimental fluctuations.}
\begin{tabular}{crr}
	\hline
	Fidelity & $F(\op{\rho}, \op{\sigma}) = $ & $0.935 \pm 0.015$ \\
	Trace distance & $D(\op{\rho}, \op{\sigma}) = $ & $0.081 \pm 0.019$ \\
	Linear entropy & $H_{lin}(\op{\rho}) = $ & $0.121 \pm 0.026$ \\
	Von Neumann entropy & $H_{vn}(\op{\rho}) = $ & $0.437 \pm 0.063$ \\
	\multirow{2}{*}{Conditional entropy} & $H_{c}(\op{\rho}|s) = $ & $-1.557 \pm 0.067$ \\
	 & $H_{c}(\op{\rho}|i) = $ & $-1.557 \pm 0.066$ \\
	\hline
\end{tabular}
\end{table}

\section*{Discussion}

To evaluate the usefulness of the four-dimensional entanglement quantitatively,
we considered the Devetak-Winter rate,
which gives the available secure key rate in a quantum key distribution against a collective attack \cite{Devetak2004}.
We assume that
Alice and Bob share the mixed state $\op{\rho}^{AB}$ and
an eavesdropper, Eve, has ancilla states with which we obtain a pure state $\ket{\psi^{ABE}}$ s.t. $\op{\rho}^{AB} = \mathrm{Tr}_E \ket{\psi^{ABE}}\bra{\psi^{ABE}}$.
In this situation,
we can define coherent information as $I_c (A \rangle B) = H_{vn}(\op{\rho}^B) - H_{vn}(\op{\rho}^{AB})$ \cite{Schumacher1996, Nielsen2010, Devetak2004}.
Therefore,
the amount of coherent information is the same as the conditional entropy when the sign is inverted.
According to ref. \cite{Devetak2004},
there exists a protocol which gives a secure key rate equal to the amount of coherent information.
In fact,
this secret key rate agrees with the key rate when we use a $d$-dimensional entangled state and $d+1$ mutually unbiased bases (MUBs) \cite{Sheridan2010}.
Therefore,
the four-dimensional entanglement in our experiment can be used as a resource for up to 1.557 bits of secure keys.
Although detailed analyses on the finite-key effect, quantum bit error rates,
and security loopholes are still needed before we can use this as a real quantum key distribution system,
this key rate gives a secure information capacity---an upper bound of the secure key rate
with an ideal measurement setup.
To use this resource for quantum key distribution,
we need to implement at least two MUBs.
One of the MUBs with respect to the computational basis $\{\ket{k}\}$ is the Fourier transform basis.
Recently,
implementations of the Fourier transform basis for a four-dimensional time-bin state have been demonstrated \cite{Ikuta2016a, Islam2017},
where cascaded MZIs were employed.
Therefore,
our experimental setup can also be used for quantum key distribution with two MUBs.
Furthermore,
it was recently pointed out that the amount of high-dimensional quantum entanglement can be bounded by measurement results using two MUBs \cite{Erker2017}.
If the amount of entanglement is only the quantity that we are interested in,
we can implement such MUBs by optimizing the MZI phases for the scheme,
which would help simplifying high-dimensional quantum communication systems.
On the other hand,
it is necessary to prepare $d + 1$ MUBs to realize the full potential of high-dimensional entanglement.
As long as a $d$-dimensional space has $d+1$ MUBs,
we can implement $d + 1$ MUBs for a time-bin qudit in principle.
For example,
a multi-arm interferometer using $d$ optical delay lines and $d - 1$ optical phase shifters,
which was used to test the high-dimensional Bell-type inequality \cite{Thew2004},
can be used to implement $d+1$ MUBs.
Although a practical implementation of $d + 1$ MUBs for a time-bin qudit remains as an important task,
our observation of four-dimensional entanglement with more than 1 bit of coherent information
 constitutes an important step towards advanced secure and dense quantum communications over a long distance.

\section*{Methods}
\subsection*{Measurement operators of the MZIs}

Here, we briefly derive the measurement operator $\op{E}_{t x \theta_1 \theta_2}$.
We assume that the expected value of photon counts is given by $n_{t x \theta_1 \theta_2} = N \mathrm{Tr} ( \op{\varrho} \op{E}_{t x \theta_1 \theta_2} )$
when a single photon in state $\op{\varrho}$ is measured $N$ times
(The details are described in ref. \cite{Ikuta2017}).
When a photon in time-bin state $\ket{k}$ enters the 2-bit delay MZI,
the state of the photon at the output port connected to the 1-bit delay MZI is $\op{M}_{1, \theta_2}^{\twobit} \ket{k}$,
where $\op{M}_{1, \theta_2}^{\twobit}$ is given by
\begin{equation}
\op{M}_{1, \theta_2}^{\twobit} = \frac{\sum_{k' = 0}^3 \left( \ket{k'} + \sqrt{\eta_1^{\twobit}} {\rm e}^{i \theta_2} \ket{k'+2} \right) \bra{k'}}{\sqrt{2 (1 + \eta_1^{\twobit})}} .
\end{equation}
Here, $\eta_1^{\twobit}$ is the transmittance ratio between the short and long paths in the 2-bit delay MZI.
Similarly,
we can define measurement operators for the 2-bit and 1-bit delay MZI at the output port connected to SNSPD 2 and 1, respectively,
as follows:
\begin{eqnarray}
\op{M}_{2, \theta_2}^{\twobit} &=& \frac{\sum_{k' = 0}^3 \left( - \ket{k'} + \sqrt{\eta_2^{\twobit}} {\rm e}^{i \theta_2} \ket{k'+2} \right) \bra{k'}}{\sqrt{2 (1 + \eta_2^{\twobit})}} , \\
\op{M}_{1, \theta_1}^{\onebit} &=& \frac{\sum_{k' = 0}^5 \left( \ket{k'} + \sqrt{\eta_1^{\onebit}} {\rm e}^{i \theta_1} \ket{k'+1} \right) \bra{k'}}{\sqrt{2 (1 + \eta_1^{\onebit})}} .
\end{eqnarray}
The post-selection at detection time slot $t$ corresponds to projection measurement $\op{P}_t = \ket{t}\bra{t}$.
From these measurement operators,
we can define $\op{E}_{t x \theta_1 \theta_2}$ as follows:
\begin{eqnarray}
\op{E}_{t 1 \theta_1 \theta_2} &=& \eta \op{M}_{1, \theta_2}^{\twobit \dag}
	\op{M}_{1, \theta_1}^{\onebit \dag}
	\op{P}_t^\dag
	\op{P}_t
	\op{M}_{1, \theta_1}^{\onebit}
	\op{M}_{1, \theta_2}^{\twobit}	,	\\
\op{E}_{t 2 \theta_1 \theta_2} &=& \op{M}_{2, \theta_2}^{\twobit \dag}
	\op{P}_t^\dag
	\op{P}_t
	\op{M}_{2, \theta_2}^{\twobit}	,
\end{eqnarray}
where $\eta$ is another transmittance ratio that compensates for differences depending on the transmittances of the optical paths and detection efficiencies of the SNSPDs.
The measurement operators for the coincidence counts are obtained by combining $\op{E}_{t x \theta_1 \theta_2}$ for Alice and Bob
and used to perform QST for two photons \cite{James2001,Thew2002}.

The transmittance ratios $\eta_1^{\twobit}$, $\eta_2^{\twobit}$, and $\eta_1^{\onebit}$ were stable
because the MZIs were fabricated by planar lightwave circuit technology.
From the previous measurement \cite{Ikuta2017},
$\eta_1^{\twobit}$, $\eta_2^{\twobit}$ and $\eta_1^{\onebit}$ for Alice (Bob) were estimated to be 1.009, 0.8300 and 1.063
(0.8495, 0.8302 and 0.9669), respectively.
On the other hand,
$\eta$ depends on the conditions of the SNSPDs.
We estimated $\eta$ from the average single photon count rates during QST
to be 0.8507 for Alice and 0.4812 for Bob.

\subsection*{Figures of merit for the entanglement}
To characterize the measured four-dimensional entanglement,
we used
fidelity $F(\op{\rho}, \op{\sigma})$,
trace distance $D(\op{\rho}, \op{\sigma})$,
linear entropy $H_{lin}(\op{\rho})$,
von Neumann entropy $H_{vn}(\op{\rho})$,
and conditional entropy $H_{c}(\op{\rho}|X)$.
Here we employed the following definitions:
\begin{eqnarray}
F(\op{\rho}, \op{\sigma}) &=& \left[ \mathrm{Tr} \sqrt{\sqrt{\op{\sigma}} \op{\rho} \sqrt{\op{\sigma}}} \right]^2,	\\
D(\op{\rho}, \op{\sigma}) &=& \frac{1}{2} \mathrm{Tr} \sqrt{( \op{\rho} - \op{\sigma} )^2},	\\
H_{lin}(\op{\rho}) &=& 1 - \mathrm{Tr} (\op{\rho}^2),	\\
H_{vn}(\op{\rho}) &=& - \mathrm{Tr} (\op{\rho} \log \op{\rho}),	\\
H_{c}(\op{\rho}|X) &=& H_{vn}(\op{\rho}) - H_{vn}(\op{\rho}^X), \\
\op{\sigma} &=& \ket{\phi} \bra{\phi},
\end{eqnarray}
where
$\op{\rho}$ is the reconstructed density operator,
$\ket{\phi}$ is given by $2^{-1} \sum_{k=0}^3 \exp(ik\phi) \ket{k}_s \otimes \ket{k}_i$,
$X \in \{s, i\}$ denotes the signal and idler photons,
and $\op{\rho}^X$ is the reduced density operator for $X$.
Since the pump pulses for SPDC were generated from continuous-wave light
and we calibrated the initial phase settings of the MZIs for Alice and Bob to maximize the extinction ratio for 1555- and 1547-nm continuous-wave light, respectively,
the generated entangled state had non-zero relative phases like $\ket{\phi}$ \cite{Ikuta2017}.
Therefore,
we optimized the phase constant $\phi$ to maximize $F(\op{\rho}, \op{\sigma})$ or minimize $D(\op{\rho}, \op{\sigma})$
as we calculated these quantities.

Conditional entropy is always positive if state $\rho$ is separable \cite{Horodecki1996,Horodecki1996a}.
Furthermore,
the minimum value of conditional entropy for a $d$-dimensional two-photon state is
$-\log_2 d$ because von Neumann entropy is always positive for any state and the maximum von Neumann entropy $H_{vn}(\op{\rho}^X)$ is $\log_2 d$.
Therefore,
a conditional entropy smaller than $-1$ bit implies that
the reconstructed state is not separable and not two-dimensional.

\section*{Acknowledgements}
We thank T. Inagaki, T. Honjo and K. Azuma for fruitful discussions.

\section*{Author contributions}
T. I. and H. T. conceived and designed the experiment and wrote the paper.
T. I. performed the experiment and data analysis.

\section*{Competing financial interests}
The authors declare no competing financial interests.

\section*{Data availability}
The data that support the findings of this study are available from the corresponding author upon reasonable request.

\bibliography{CitationListForQuditsTrans}

\end{document}